\documentclass[11pt,twoside]{article}
\usepackage{asp2006}
\usepackage{epsf}
\usepackage{graphics}
\usepackage{lscape}
\def\edcomment#1{\iffalse\marginpar{\raggedright\sl#1\/}\else\relax\fi}
\reversemarginpar

\begin{document}

\title{On the Relationship Between a CME Associated Waves Observed on 5 March 2000}
\author{N.-E. Raouafi$^1$ and D. Tripathi$^2$}
\affil{$^1$National Solar Observatory, 950 N. Cherry Avenue, Tucson, AZ 85719, USA}
\affil{$^2$Department of Applied Mathematics and Theoretical Physics, University of Cambridge,
Wilberforce Road CB3 OWA, UK}

\begin{abstract} We study the relationship between different wave phenomena associated with a
coronal mass ejection (CME) observed on  05 Mar. 2000. EIT waves were observed in the images
recorded by EIT at 195~{\AA}. The white-light LASCO/C2 images show clear deflection and propagation
of a kink along with the CME. Spectroscopic observations recorded by the UVCS reveals excessive
line broadening in the two O~{\sc{vi}} lines (1032 and 1037 {\AA}). Moreover very hot lines such as
Si ~{\sc{xii}} and Mg ~{\sc{x}} were observed. Interestingly, the EIT wave, the streamer deflection
and the intensity modulation along the slit were all propagating North-East. Spatial and temporal
correlations show that the streamer deflection and spectral line broadening are highly likely to be
due to a CME-driven shock wave and that the EIT wave is the signature of a CME-driven shock wave in
the lower corona. \end{abstract}

\vspace{-0.5cm}
\section{Introduction}

Coronal mass ejections (CMEs) are one of the most fascinating and intriguing forms of solar
activity. They occur when solar plasma threaded with topologically complex magnetic fields are
ejected out into the corona and interplanetary medium. Recent technological developments have
yielded a steady increase in qualitative as well as quantitative studies of CMEs and related
phenomena. However, numerous CME-associated phenomena such as the relationship between different
wave features (EIT waves: Thompson et al. 1998; CME-driven shock waves: Hundhausen et al. 1987) are
not understood unambiguously. CMEs with speed greater than the Alfv\'en speed of the local plasma
produce shock waves whose effects can be traced through radio type II bursts (Klassen et al. 2000)
and/or emission of very hot lines (compared to the ambient coronal temperatures; Raouafi et al.
2004). These waves could also be detected through deflections in remote streamer belts with respect
to the ejected CME material (see Hundhausen et al. 1987 and Sime \& Hundhausen 1987). The high
quality data of SOHO/LASCO (Brueckner et al. 1995) yielded numerous cases for streamer deflections
associated with super-Alfv\'enic CME eruptions (see Sheeley et al. 2000). These observations led to
the conclusion deduced by the previous authors that these deflections were indeed a consequence of
CME-driven shock waves.

Spectroscopic signatures of coronal shock waves due to high speed CMEs ($>600$~km~s$^{-1}$; see
Raymond et al. 2000) are line broadening (e.g., \ion{O}{vi} 1032~{\AA} \& 1037~{\AA} lines) and
enhanced emissions in spectral lines that are rarely observed in the corona such as \ion{Si}{xii}
520~{\AA} and \ion{Mg}{x} 625~{\AA} (e.g., Raouafi et al. 2004) due to their relatively high
formation temperatures ($>2$~MK). Up to now only a few CME-driven shock wave events have been
reported through SOHO/UVCS (Kohl et al. 1995) observations: Raymond et al. 2000; Mancuso et al.
2002; Raouafi et al. 2004; and Ciaravella et al. 2005 (hereafter RMRC00-05).

Since the discovery of the H${\alpha}$ Moreton waves (Moreton 1964), it was thought that these
waves were due to the intersection of coronal shock waves (due to flares) with the chromosphere
(e.g., Uchida 1968). Later when EIT waves (Thompson et al. 1998) were discovered based on the
observations recorded by EIT (Delaboudini\`ere et al. 1995), they were interpreted as the coronal
manifestation of the chromospheric Moreton wave (Thompson et al. 1999). However, the difference in
propagation patterns and speeds of Moreton and EIT waves questions the similarity between these two
phenomena.

Although these waves are being observed more frequently the question remains open as to how these
different wave phenomena are related to each other. Based on MHD simulations, (Chen et al. 2005a,b)
showed that Moreton waves are the surface counter part of the CME driven shock wave. However, the
EIT waves are the slow moving wave fronts traveling behind the Moreton wave due to the opening of
the magnetic flux system. This was also suggested by Zhukov \& Auch\`ere (2004) and by Delanee
(2000). However, this relationship and the nature of EIT waves remains elusive.

A CME event on 5 March 2000 has been observed simultaneously by different instruments on board SOHO
(EIT, LASCO and UVCS) with different signatures of wave features. This event provides a unique
opportunity to establish a relationship between different phenomena such as the CME driven shock
wave, streamer deflection and the EIT waves. Observations and data analysis are described in
\S2 and results and conclusions are presented in \S3.

\section{Observations, data analysis and results \label{obs}}
\subsection{LASCO Observation of the CME and streamer deflection}

 \begin{figure}[!h]
 \plotone{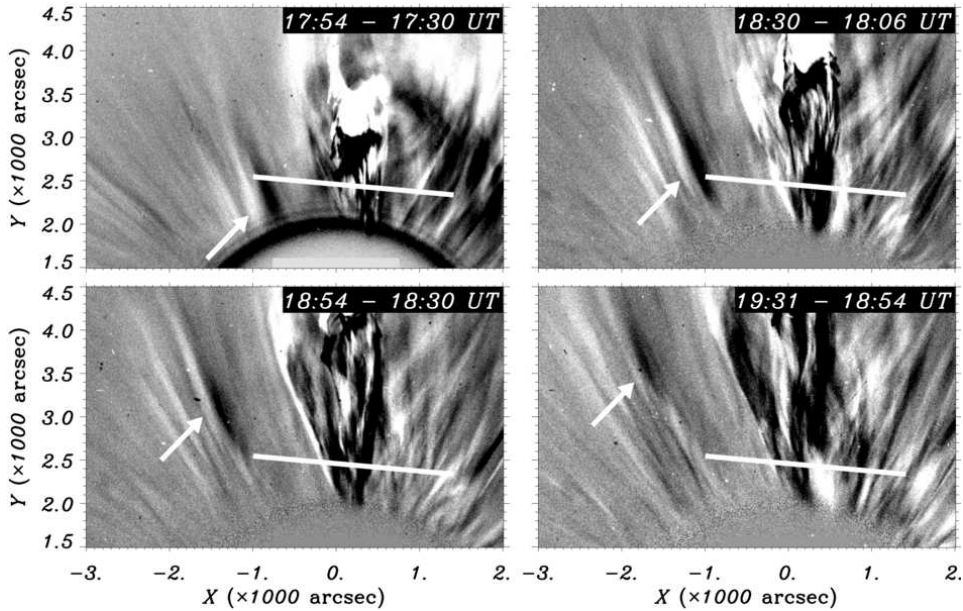} 
 \caption{Running difference of LASCO-C2 white-light images. The arrows locate the propagating kink in
     the streamer. The location of the UVCS slit is shown by white-straight line. \label{lasco}}
 \end{figure}

The CME event of 5 March 2000 appeared as a three-part structure (bright front followed by a dark
cavity and a bright core) in the LASCO/C2 field of view (FOV) at 16:54 UT. Images with FOV
extending from 2.5-6.0~R$_{\sun}$ were recorded every 20 minutes. Fig.~\ref{lasco} displays the
running difference of LASCO/C2 images of the white-light CME showing the proper motion of features
(Sheeley et al. 1999). The bright material seen as a core of the CME is the erupting prominence
which had expanded. The leading edge of the CME propagated with a super Alfvenic speed of 860 km/s
(Tripathi et al. 2006).

The arrows mark the location of the streamer deflection propagating North-East. As depicted in
Fig.~\ref{lasco}, there is no evidence of plasma material which could have caused such a kink in
the streamer. The simultaneous propagation of the kink in the streamer outward in the corona along
with the CME provides strong evidence for the existence of a wave phenomenon traveling in the flank
of the CME. The speed of the propagating kink (projected on the plane of the sky), measured by
tracking the boundary between the bright and dark features from the running difference images, was
about 260~km~s$^{-1}$. Sheeley et al. (2000) found that speed of the CME-flank material is likely
to be smaller than that of the bright leading edge (860~km~s$^{-1}$ in the present case). Since the
CME propagates with super-Alfv\'enic speed, it can drive a shock. Therefore it is plausible to
conclude that the propagating kink was likely to be the consequence of shock wave driven by the
associated CME.

\subsection{UVCS observations}

The UVCS observational sequence started at 15:58~UT and ended at 20:12~UT. This time interval
covers adequately the time evolution of the CME event that first appeared in the LASCO at around
16:54~UT. The 40 arcmin long UVCS slit was centered at 2.55~R$_{\sun}$ from Sun center at
$355^\circ$ counterclockwise (CCW) from the north pole. 80 exposures of 180 seconds each have been
recorded in the wavelength ranges of 1027-1042~{\AA} and 1241-1253~{\AA} for the main and
redundant  channels, respectively. The latter includes the \ion{Mg}{x} 625~{\AA} line (observed in
second order), where the former contains the strong \ion{O}{vi} doublet 1032~{\AA} \& 1037~{\AA}
together with other weaker lines such as \ion{Si}{xii} 520~{\AA} which is observed in second order.
However, we concentrate on the first 21 exposures where hot plasma emission is observed before the
CME material has reached the slit.

\begin{figure}[!h]
 \plotone{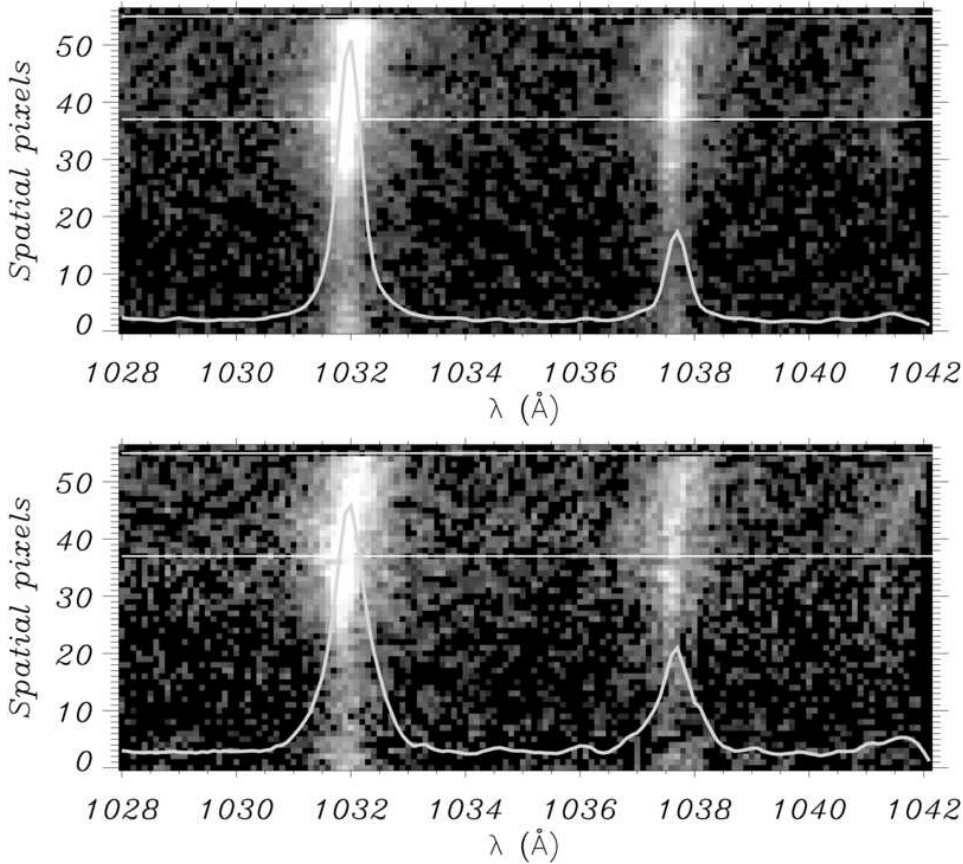} 
\caption{Averages of exposures 1-15 (top) and 16-21 (bottom) of the UVCS observation
sequence of 5 March 2000 at 15:58~UT. The spectral lines shown are the
\ion{O}{vi} doublet at 1032~{\AA} and 1037~{\AA} and the weak line of
\ion{Si}{xii} at 520~{\AA} observed in second order. The spectra
overplotted on each panel are the corresponding spatially-binned ones
between the two horizontal lines. The focus is on the shape of the
profiles and then the amplitudes of the profiles are indirectly
proportional to the real ones. \label{March052000_12072006_OVI}}
\end{figure}

\begin{figure}[!h]
\plotone{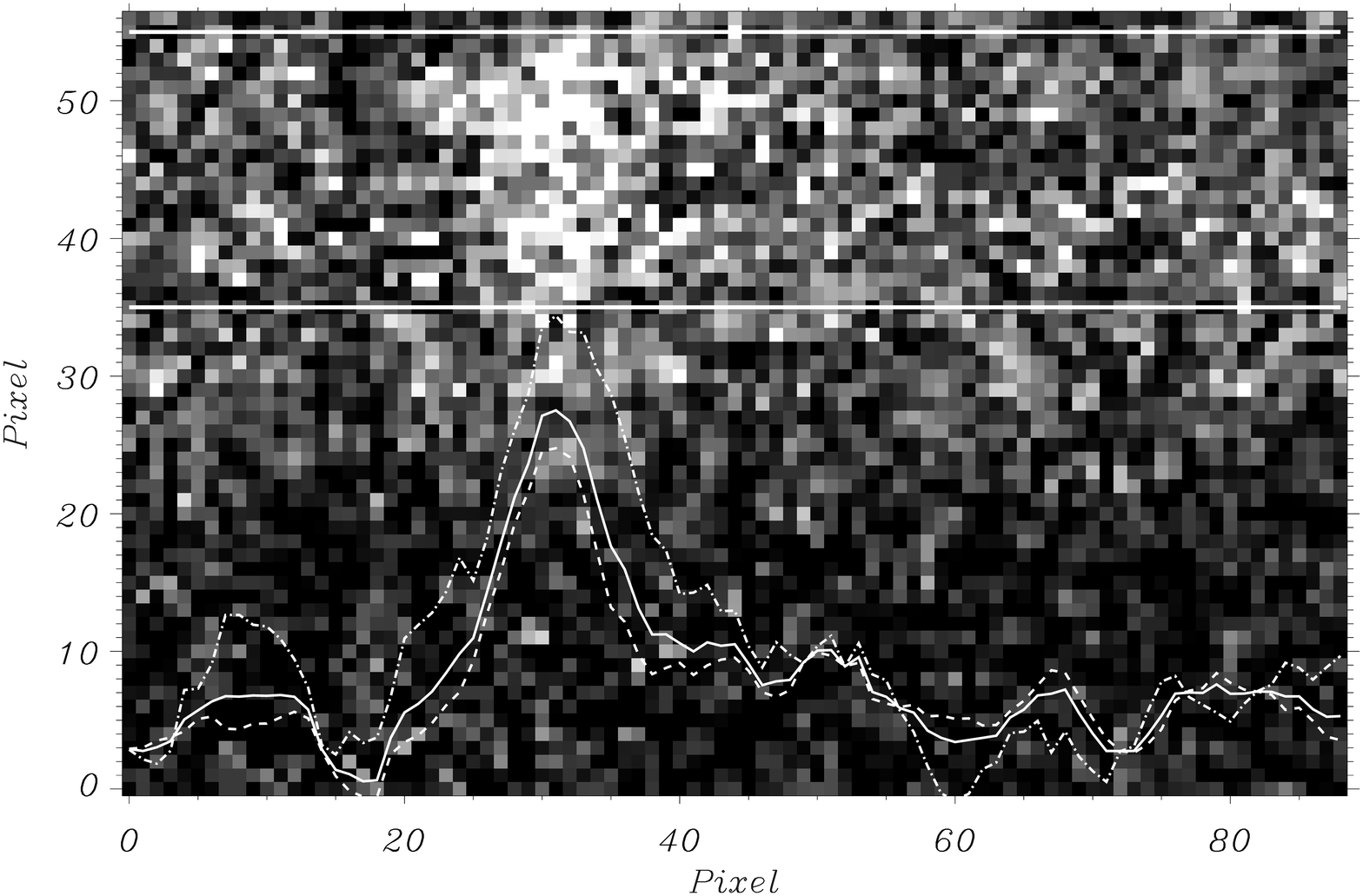}
\caption{Average of the first 21 exposures recorded in the redundant channel of the UVCS telescope
showing an intensity enhancement of the \ion{Mg}{x} 625~{\AA} line. The normalized spectra averaged
over spatial bins between the two horizontal lines are overplotted (solid: exposures 1--21; dashes:
1--15; dot-dashes: 16--21). Note the intensity enhancement and the excessive broadening of the
\ion{Mg}{x} line in the exposures 16--21, which occurs simultaneously with the broadening and
intensity enhancement of the \ion{O}{vi} and \ion{Si}{xii} lines (see
Fig.~\ref{March052000_12072006_OVI}). \label{March052000_event_MgX}}
\end{figure}

Fig.~\ref{March052000_12072006_OVI} displays two sets of averaged UVCS exposures (top panel:
average of exposures from 1 to 15; bottom panel: average of exposures from 16 to 21) showing the
\ion{O}{vi} doublet together with the \ion{Si}{xii} 520~{\AA} line. The overplotted spectra are
spatially binned between the two horizontal lines plotted on the figure. Note that the amplitude of
the spectral lines are not on scale and only the change in the profile shapes are emphasized. The
\ion{O}{vi} line profiles in the top panel are composed mainly of a central narrow component and
emission in the \ion{Si}{xii} line is weak but remarkable. In the bottom panel, the \ion{O}{vi}
lines are dimmed compared to the previous one and are significantly wider as shown by the
overplotted spectra. The emission profiles in the \ion{Si}{xii} line are enhanced and are also
wider.

Fig.~\ref{March052000_event_MgX} shows the average of the first 21 exposures in the redundant
channel covering the \ion{Mg}{x} 625~{\AA} (second order). The spatially binned spectra between the
horizontal lines are overplotted. The solid curve is the spatially binned spectra of all the 21
exposures. The significance of the other curves is explained in the caption of the same figure. The
dot-dashed spectrum (average of exposures 16--21) shows an intensity enhancement and profile
broadening of the \ion{Mg}{x} line simultaneously with the broadening of the \ion{O}{vi} and
\ion{Si}{xii} lines.

The increase in the line widths together with enhanced emissions in hot lines are very likely due
to plasma heating which is very likely due to a shock wave propagating in front of the CME (e.g.,
Raouafi et al. 2004 and Ciaravella et al. 2005). A simple time computation suggested that the shock
wave has reached the UVCS slit after 16:45 and the heated gas emission lasted till about 17:20 UT,
where emission in colder lines is observed. Extra broadening in the \ion{O}{vi} lines is noticeable
even after the CME cold material has reached UVCS FOV. The intensity modulation along the slit, in
particular the upper section of the slit, shows structures drifting towards the upper end of the
slit, which corresponds to the North-East direction (see Tripathi \& Raouafi 2007). In addition,
intensity dimming in the upper section of the slit of the \ion{O}{vi} 1032~{\AA} line from one
exposure to the next, where that of the 1037~{\AA} line changes quite differently. It gets enhanced
in particular in exposure number 20. This is evidence for the acceleration of the \ion{O}{vi} ions
that leads the 1032~{\AA} line to run out of resonance and the 1037~{\AA} line to get optically
pumped by the chromospheric lines of \ion{C}{ii}.We believe that is evidence that the physical
process causing the intensity variation and lines' broadening (as shown previously) is moving along
the slit to the North-East quadrant of the solar corona.

\subsection{EIT observations: EIT Wave}

\begin{figure}[!h]
\plotone{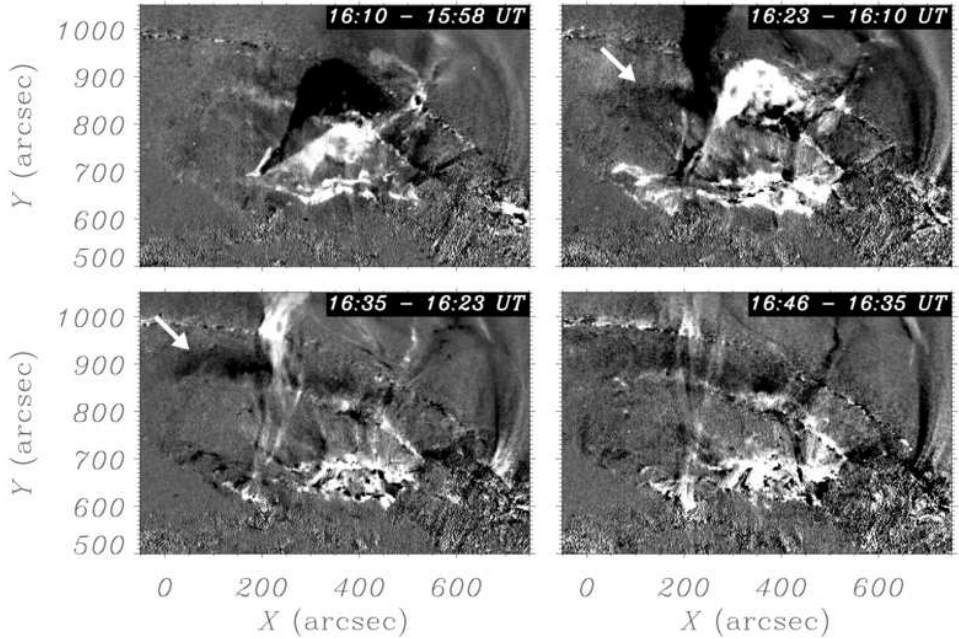} 
\caption{EIT running difference images showing the EIT waves. Arrows
locate the EIT wave front. \label{ewave_img}}
\end{figure}

SOHO/EIT images at 195~{\AA} are recorded regularly with a cadence of 12 mins and were used for the
present study. Fig.~\ref{ewave_img} displays the running difference images taken by EIT at
195~{\AA}. The filament slowly rose and erupted at 15:46~UT. The EIT wave associated with this
eruption first appeared at 16:10~UT and disappeared at 16:58~UT.

The EIT wave is seen propagating only towards the North-East. No counter-part moving to the
South-West was detected probably because of the presence of an active region. The propagating
wave-front also did not traverse through the coronal hole which was located North-East of the
source region, similar to the observations presented by Thompson et al. (1999). The estimated
average speed of the propagation was about 55~km~s$^{-1}$. Note that no correction for the
projection effect was performed while measuring the speed. Except for the speed being much lower,
other characteristics are similar to those of the EIT waves. 

\section{Summary of results and discussion}

We studied the relationship between different coronal waves (EIT wave, streamer deflection and a
CME-driven shock wave) generated by the CME event on 5 March 2000 as observed by different
instruments on board SOHO (EIT, LASCO and UVCS).

LASCO-C2 images show a deflection propagating outward at $\sim260$~km~s$^{-1}$ (projected on the
plane of the sky) in a remote streamer located approximately at $10^\circ-15^\circ$ CCW from north
pole. The streamer kink was first seen in LASCO-C2 FOV ($\sim2.5$~R$_{\sun}$) at about 17:30~UT. No
evidence for any CME material which could be the origin of such a deflection. The speed of the
leading edge of the CME was $\sim860$~km~s$^{-1}$, which is sufficient to generate a shock wave.
Therefore, this propagating kink in the streamer provides strong evidence for a CME-driven shock
wave in the corona as suggested by Sheeley et al. (2000).

UVCS spectra show excessive broadening in \ion{O}{vi} lines and intensity enhancement in the hot
lines of \ion{Si}{xii} 520~{\AA} and \ion{Mg}{x} 625~{\AA}. These are clear evidence for a
CME-driven shock wave (see RMRC00-05). The analysis of the intensity modulation along the slit also
reveals the propagation direction of the wave, North-East, that is the same as the kinking
streamer. UVCS observations show that the shock wave would have reached the UVCS slit around
16:45~UT. 

EIT difference images show evidence for an EIT wave front propagating North-East with speed of
$\sim55$~km~s$^{-1}$ (the solar disk shape not taken into account) up to the north polar hole where
the propagation has stopped, satisfying the properties of EIT waves. The propagation of the wave in
the South-East direction seems to be disabled by the presence of the complex active region. The EIT
wave first appeared at 16:10~UT and disappeared gradually after 16:58~UT.

On one hand, temporal and spatial correlations between the events observed by UVCS and LASCO
suggest that they are basically different manifestations of the same phenomenon. Measurements of
the speed in different part of CME envelopes reveal that noses of CMEs travel faster than their
flanks (Sheeley et al. 2000). Therefore we anticipate that the propagating shock would have a
similar property and would have speed higher in front of the CMEs and lower in the flanks.
Different speed in the different parts would provide evidence for the anisotropic propagation of
associated shock waves. On the other hand, the EIT wave observed in the low corona was
significantly slower the UVCS and LASCO events. A possible explanation for this is the opening of
the magnetic flux system due to the expulsion of the CME rather than the manifestations of the CME
driven shock wave as suggested by Chen et al. (2005a,b) based on MHD modeling and by Delanee (2000)
and by Zhukov \& Auch\`ere (2004). Moreton waves, which are observed in the chromosphere are
interpreted as counterpart of CME driven shock waves (e.g., Chen et al. 2005a). Unfortunately, we
did not have H${\alpha}$ high cadence observations for this event, which prevented a comparison
with a chromospheric Moreton wave. However, the semi-circular wavefront was moving in the same
direction as the coronal shock wave. This supports the hypothesis that all three features are
linked to each other. They are generated by the same source, propagate in the same direction and
have good time and spatial correlations. The only apparent problem resides in the propagation
speed. However, note that the three wave-like structures do not share similar physical conditions.
One of these is the space and its characteristics. The speed of Alfv\'en waves is given by
$\displaystyle{V_A=\frac{B}{\sqrt{4\pi\rho}}}$, where $B$ is the magnetic field strength and $\rho$
is the density. Depending on models, the density drops by 3 to 4 orders of magnitude or more
between the very sparse corona and 2.0-3.0~R$_{\sun}$, where the magnetic field drops by an order
of magnitude or more (assuming the the coronal field is nearly potential and thus drops as
$\displaystyle{r^{-2}}$; see Altschuler \& Newkirk 1969). While the high corona is increasingly
less dense with increasing altitude the corresponding Alfv\'en speed tends to increase. This is not
the case in the lower corona where the plasma density is higher and thus is characterized by a
smaller Alfv\'en speed. This may explain at least partially the difference in speed. Another
parameter is the propagation direction of the waves. The CME-driven shock wave is mainly
propagating parallel to the magnetic field lines. However, the wave propagating on the solar disk
does not share this property and propagates across. It encounters different magnetic structures
that may allow for energy leakage which might consequently damp and slow down the wave (see for
instance De Pontieu et al. 2004 on energy leakage of the acoustic oscillation $p$ modes through
flux tubes with high inclinations to the chromosphere which contributes to the heating of this
layer). A good illustration for that is given by active regions through which EIT waves could not
propagate and also across polar holes. We think that energy carried by the wave is progressively
leaked to different magnetic structures which may heat and accelerate the plasma in these
structures. This interpretation needs to be carried further and deeper through additional analysis
of other observational examples and also by numerical simulations.

\acknowledgements
The National Solar Observatory is operated by the Association of Universities for Research in
Astronomy, Inc., under cooperative agreement with the National Science Foundation. NER's work
is supported by NSO and NASA grant NNH05AA12I. DT acknowledges the support from STFC. SOHO is a
project of international collaboration between ESA and NASA.

\end{document}